\documentclass[conference]{IEEEtran}
\IEEEoverridecommandlockouts
\usepackage{cite}
\usepackage{amsmath,amssymb,amsfonts}
\usepackage{algorithmic}
\usepackage{graphicx}
\usepackage{textcomp}
\usepackage{xcolor}
\usepackage{multirow}
\usepackage{algorithm}
\usepackage{amsmath}
\usepackage{booktabs}
\usepackage{colortbl}
\usepackage[hyphens]{url}
\usepackage[hidelinks]{hyperref}
\usepackage{soul}
\usepackage{subcaption}
\hypersetup{breaklinks=true}

\def\BibTeX{{\rm B\kern-.05em{\sc i\kern-.025em b}\kern-.08em
    T\kern-.1667em\lower.7ex\hbox{E}\kern-.125emX}}

\newcommand{\wh}[1]{\textsc{Whisper}}
\newcommand{\nm}[1]{\textsc{Nemo}}
\newcommand{\tie}[1]{\textbf{\textsc{TIE}}}
\newcommand{\vox}[1]{\textbf{\textsc{Vox-Populi}}}
\newcommand{\ted}[1]{\textbf{\textsc{TedLium}}}
\newcommand{\fle}[1]{\textbf{\textsc{Fleurs}}}
\newcommand{\dem}[1]{\textsc{DEmucs}}
\newcommand{\li}[1]{\textsc{Le}}
\newcommand{\sg}[1]{\textsc{Sg}}
\newcommand{\mg}[1]{\textsc{MetricGAN+}}
\newcommand{\den}[1]{$\mathcal{DENOASR}$}

\begin{document}

\title{\den{}: Debiasing ASRs\\ through Selective Denoising
}

\author{
\IEEEauthorblockN{Anand Kumar Rai}
\IEEEauthorblockA{
\textit{IIT Kharagpur, India} \\ 
\textit{Joint Plant Committee, India}}
\and
\IEEEauthorblockN{Siddharth D Jaiswal}
\IEEEauthorblockA{
\textit{IIT Kharagpur, India}}
\and
\IEEEauthorblockN{Shubham Prakash}
\IEEEauthorblockA{
\textit{IIT Kharagpur, India}}
\and
\IEEEauthorblockN{Bendi Pragnya Sree}
\IEEEauthorblockA{
\textit{IIT Kharagpur, India}}
\and
\IEEEauthorblockN{Animesh Mukherjee}
\IEEEauthorblockA{
\textit{IIT Kharagpur, India}}
}

\maketitle

\begin{abstract}
 Automatic Speech Recognition (ASR) systems have been examined and shown to exhibit biases toward particular groups of individuals, influenced by factors such as demographic traits, accents, and speech styles. Noise can disproportionately impact speakers with certain accents, dialects, or speaking styles, leading to biased error rates. In this work we introduce a novel framework \den{} which is a selective denoising technique to reduce the disparity in the word error rates between the two gender groups \textit{male} and \textit{female}. We find that a combination of two popular speech denoising techniques viz. \dem{} and \li{} can be effectively used to mitigate ASR disparity without compromising their overall performance. Experiments using two SOTA open-source ASRs -- OpenAI \wh{} and NVIDIA \nm{} on multiple benchmark datasets -- \tie{}, \vox{}, \ted{} and \fle{} show that there is a promising reduction in the average word error rate gap across the two gender groups. For a given dataset, the denoising is selectively applied on speech samples having speech intelligibility below a certain threshold estimated using a small validation sample thus ameliorating the need for large-scale human written ground-truth transcripts. Our findings suggest that selective denoising can be an elegant approach to mitigate biases in present-day ASR systems. 
 \end{abstract}


\section{Introduction}
In recent years, automatic speech recognition (ASR) systems have witnessed significant advancements in terms of accuracy and efficiency \cite{garnerin2019gender}, and are now used for various applications such as virtual assistants~\cite{Siri_Live,alexa,Cortana_Live}, transcription services~\cite{Youtube_Live,Zoom_Live,Bing_Speech}, and voice-controlled devices~\cite{dong2020secure}. 
Despite these remarkable strides, even state-of-the-art ASRs are unable to provide equitable performances across various demographic traits like gender~\cite{tatman2017effects,tatman2017gender,feng2021quantifying}, age~\cite{dichristofano2022performance,vipperla2010ageing,feng2021quantifying}, race~\cite{koenecke2020racial}, speech impairment~\cite{tu2016relationship,shahamiri2021speech} and non-native accent~\cite{dichristofano2022performance,feng2021quantifying}. These disparities prevent wide-scale adoption of ASRs~\cite{koenecke2020racial,ngueajio2022hey}, specially amongst the discriminated groups.

Various studies \cite{kumalija2022performance,dua2023noise} have highlighted the detrimental impact of noise on the overall performance of ASR systems, resulting in errorneous transcriptions. These studies underscore the critical need for noise-robust ASR systems to improve performance. The impact of noise varies across different speech styles~\cite{sadeghi2018effect}, potentially contributing to performance disparities. In this work, we have attempted to evaluate the impact of denoising on mitigating the effects of speaker demographics in ASR tools performance. 

In signal processing, denoising~\cite{burwen} is used to remove or reduce unwanted noise and preserve essential information in an audio or image sample. Specifically for audio samples, this technique can be used to improve clarity, intelligibility and the overall listening experience, particularly in scenarios with background noise or interference. Similarly, for ASRs denoising is relevant as it enhances the transcription accuracy by removing background noise and improving the clarity of the input audio.

\begin{figure}[!t]
    \centering
    \includegraphics[width=0.4\textwidth]{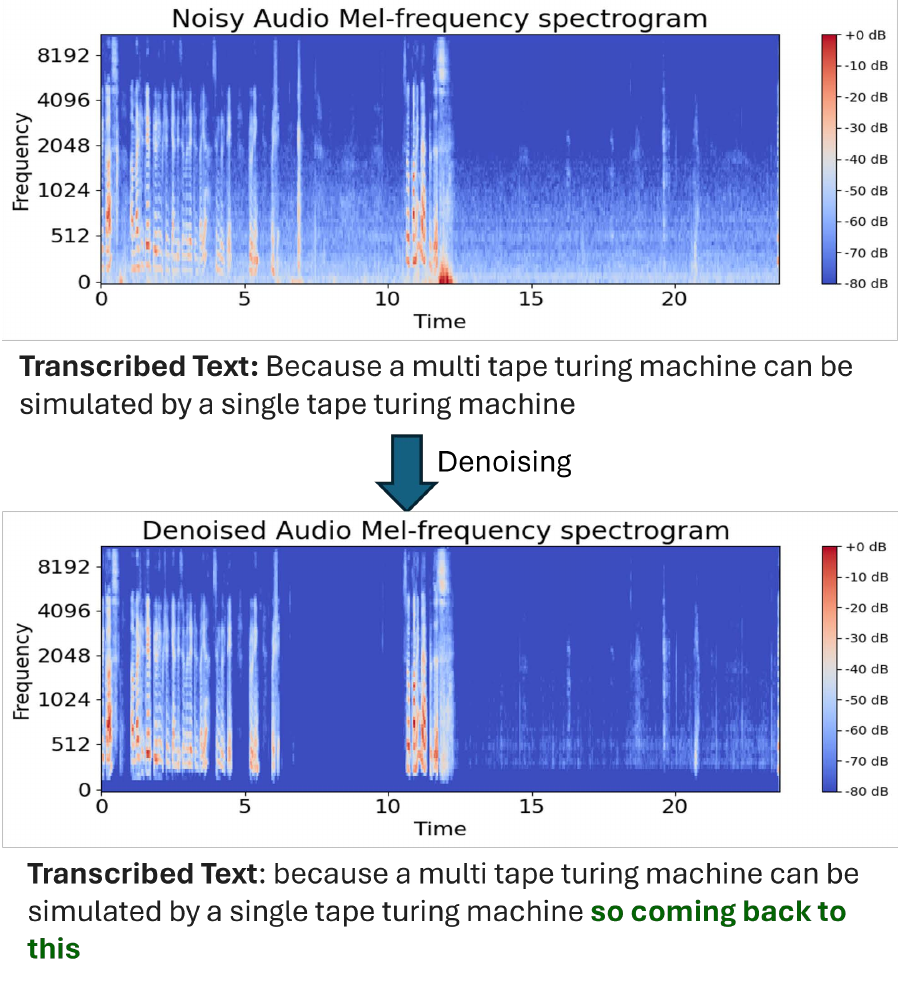}
    \caption{\footnotesize \bf Effect of denoising in speech spectogram and eventually on ASR transcription performance. The text highlighted in green gets omitted from ASR transcription in presence of noise in the spectogram. \wh{} has been used for transcribing a speech sample from the \tie{} dataset while the denoising strategy used was \dem{} followed by \li{}.}
    \label{fig:noisy}
\end{figure}
\subsection{Denoising algorithms}
Many denoising algorithms have been proposed in literature, such as spectral subtraction~\cite{boll1979suppression}, Wiener filtering~\cite{van2009speech} and, more recently, deep learning based techniques~\cite{michelsanti2021overview}. These methods estimate noise using its statistical characteristics, establish a criteria for noise distinction, and then separate the noise from the audio signals, resulting in clearer audio.
Spectral Gating~\cite{sainburg2020finding} (\sg{}) and Line Enhancement~\cite{karraz2021effect} (\li{}) are two state-of-the-art denoising algorithms based on traditional methods proposed recently in literature. \sg{} is a variant of Noise Gate~\cite{records} and utilizes reference noise to estimate the noise traits and reduce it. \li{} exploits noise and desired signal frequency distinctions in the time-domain by utilizing high-pass filters.

\dem{} (Deep Multichannel Convolutional Neural Network for Speech Enhancement)~\cite{defossez2019music} and \mg{}~\cite{fu2021metricgan+} are cutting-edge denoising algorithms leveraging deep learning techniques. \dem{} employs multichannel convolutional layers to effectively separate speech signals from noise, while \mg{} utilizes a generative adversarial network framework to generate high-quality denoised speech signals. \\
\if{0}\noindent \textbf{Denoising MOOC datasets}: TIE~\cite{rai2023deep}, a recently released large-scale dataset
has more than 8700 hours of speech data of Indian university teachers delivering technical lectures as part of the NPTEL MOOC program~\cite{krishnan2009nptel}. While examining the lecture videos in the dataset, we noticed that speaker movement, blackboard writing, student interactions, etc. during the lecture induced noise in some of the lecture recordings. 
Hence, this large-scale data serves as a perfect test bed for applying the aforementioned denoising algorithms.\\ \fi
\subsection{Denoising for ASRs} 
The presence of noise often results in addition, omission and substitution of words in the ASR generated transcripts as compared to ground-truth ones. One such example has been illustrated in Figure \ref{fig:noisy}. Earlier research works have focused on building noise-robust ASRs~\cite{li2014overview,chao2021tenet,narayanan2014joint} with noise adaptive training, and some studies have proposed employing advanced noise reduction techniques as a pre-processing strategy to enhance overall ASR performance~\cite{yadava2022enhancements}. These techniques have proven beneficial in enhancing the robustness of ASR performance when there is a mismatch in the noise characteristics between the training and test sets. However, none of these solutions have used denoising as a bias mitigation strategy, i.e., to reduce the disparity in the ASR performance across speakers with varied demographic attributes. Ours is the first work to examine the impact of denoising on debiasing ASR performance, exploring how noise affects speaker style as mentioned in \cite{sadeghi2018effect}  and, consequently, ASR accuracy.

\if{0}
\subsection{Towards Debiasing ASRs}
While addressing disparities in commercial ASRs may be challenging due to their limited accessibility in the public domain, it is still possible to evaluate and benchmark their performance using publicly available demographic annotated ASR datasets. Datasets like AAVE ~\cite{AAVE}, Arte Bias ~\cite{meyer2020artie}, VoxPopuli~\cite{wang2021voxpopuli}, and TIE ~\cite{rai2023deep} can serve as valuable resources for assessing fairness in the ASR models across demographic subgroups before their release.

To address the issue of disparity open source ASRs, ~\cite{meyer2020artie} proposed fine-tuning the pre-trained 
model on the representative dataset and then, on each target demographics one at
a time. Some of the other strategies involve data augmentation with diverse samples from underrepresented demographics or samples with added noise, reverberation and speed perturbation.Additionally, implementing bias-correction techniques during training or developing adversarial learning ore transfer learning methods ~\cite{vu2014improving,sullivan2022improving} to reduce bias for non native speakers might contribute to more equitable ASR models.  
\fi 




\if{0}
The researchers investigated the quality disparity in the Automatic Speech Recognition (ASR) generated transcripts. They measured this disparity using the median Word Error Rate (WER) differences between transcripts obtained from YouTube automatic captions and OpenAI Whisper. The study revealed discrepancies in the ASR-generated transcripts of both ASRs they evaluated based on factors such as gender, region, age and speech rate of the speakers, as well as the discipline category of the lectures.

To evaluate and address the disparities of Whisper and Wav2Vec2, we carefully sampled a test dataset from TIE dataset that is representative of the entire corpus and have used two noise-filtering techniques as a pre-processing step for fine-tuning experiments: Spectral Gating and Line Enhancement.

Spectral Gating, a variant of the Noise Gate\footnote{https://en.wikipedia.org/wiki/Noise{\_}gate}, operates in the frequency domain, utilizing reference noise to estimate noise traits and employing thresholding to preserve desired audio while reducing noise.

Spectral Gating , which is a form of Noise Gate \footnote{https://en.wikipedia.org/wiki/Noise{\_}gate}, is a noise reduction method that operates in the frequency domain, using a reference noise segment to estimate noise characteristics and thresholding to remove noise while preserving the desired audio content.

Line Enhancement, on the other hand, exploits the distinct frequency regions occupied by noise and desired signal components in the time-frequency domain. It involves passing the audio signal through a high pass filter and reducing noise-dominated regions to produce an enhanced audio signal with preserved signal quality.

\fi



\subsection{Research questions} 
We now state the research questions tackled in this study.

\noindent \textbf{RQ1.} Does denoising audio samples as a pre-processing step in the ASR pipeline impact the accuracy and enable disparity reduction across sensitive demographic attributes like gender in open-source ASRs like \wh{} and \nm{}? 

\noindent \textbf{RQ2.} Which among the denoising techniques -- traditional signal processing based or deep learning based or an amalgamation -- is the most effective preprocessing approach in reducing disparity without compromising the accuracy? 

\noindent \textbf{RQ3.} Should denoising be applied blindly to all speech samples or to a selected set of samples based on an appropriate thresholding heuristic at inference time?




\subsection{Our contributions}
In this work, we address the research questions stated above for two state-of-the-art open-source ASRs -- \wh{} and NVIDIA \nm{}.
In particular, we introduced the \den{} framework to examine the impact of denoising techniques on ASR performance and disparity reduction. Through preprocessing steps using signal processing (\sg{} \& \li{}) and deep learning-based methods (\dem{} \& \mg{}), we assessed their influence on ASR accuracy and disparity reduction across the demographic group gender. We used multiple datasets to demonstrate the effectiveness of our results, viz. \tie{}~\cite{rai2023deep}, \vox{}~\cite{wang2021voxpopuli}, \ted{}~\cite{rousseau2012ted} and \fle{}~\cite{conneau2023fleurs}.
Our findings indicate successful debiasing of both the ASRs addressing \textbf{RQ1}. Subsequently, we identified that \dem{} followed by \li{} as the optimal denoising combination for superior effectiveness in mitigating disparities without compromising accuracy and latency (addresses \textbf{ RQ2}). Lastly for addressing \textbf{RQ3}, we compared the effectiveness of inferencing on selectively denoised audio samples, based on an intelligibility threshold obtained from a small validation set, and found that it resulted in further gains. Our findings indicate that selective denoising is more effective in debiasing ASR performance. Overall we observe that this method reduces the absolute word error rate gap between male and female voice transcripts by 11\%, 100\%, 71\%, 100\% (in case of \wh{}) and 22\%, 100\%, 77\%, 21\% (in case of \nm{}) for the datasets \tie{}, \vox{}, \ted{} and \fle{} respectively. 

\section{Related Work}

Existing literature has usually had a strong separation between denoising of speech signals and debiasing of ASR systems. 

\subsection{Denosing of speech signals} 

Denoising of speech signals is a crucial area of research in both signal-based and deep learning-based methods, particularly for its impact on ASR systems. 

Denoising has been performed using traditional signal processing-based methods, such as spectral subtraction~\cite{boll1979suppression}, Wiener filtering~\cite{venkateswarlu2011improve}, and adaptive filtering~\cite{karraz2021effect}, which aim to enhance speech clarity by reducing background noise through mathematical models and signal processing algorithms~\cite{garg2016comparative,sambur1978adaptive}. These methods, while effective, often struggle with non-stationary noise and can introduce artifacts. In contrast, modern deep learning-based approaches, leveraging neural networks like CNNs~\cite{pandey2019new}, RNNs~\cite{osako2015complex}, auto-encoder based models~\cite{lu2013speech} and transformer-based models~\cite{wang2021tstnn}, have shown significant improvements in denoising performance by modeling complex noise patterns and enhancing speech signals~\cite{grozdic2017whispered,soe2020discrete,chandrakala2023denoising,lee2024knowledge}. 

These techniques are reliable and sufficient to denoise audio samples, resulting in improved ASR performance. They can be broadly classified into two groups: (a) speech enhancement of test samples using various noise filters and other deep learning-based noise reduction techniques~\cite{yu2020speech,weninger2015speech,yuliani2021speech,michelsanti2021overview}, and (b) noise adaptive training of ASR systems~\cite{narayanan2014joint,zhu2023robust}. Effective denoising is vital for ASR as it directly influences the accuracy of speech recognition systems, ensuring that transcriptions are accurate even in noisy environments. This is particularly important for applications in real-world scenarios where background noise is ubiquitous, such as in virtual assistants, telecommunication, and voice-controlled systems.

\subsection{Debiasing of speech signals} 
Although stable performance is one of the goals of an ASR platform, recent research has also focused on aspects like equitable performance across multiple speaker characteristics -- demography~\cite{dichristofano2022performance,vipperla2010ageing,feng2021quantifying}, speech rate, gender~\cite{tatman2017effects,tatman2017gender,feng2021quantifying}, etc.
 This performance parity is provided using debiasing techniques, which have two primary aspects -- (a)~augmenting models with data samples from under-represented groups~\cite{panayotov2015librispeech,garnerin2019gender,zhangcomparing,feng2021quantifying,sari2021counterfactually} and, (b)~model adaptation or fine-tuning ~\cite{meyer2020artie,winata2020learning} toward unseen accents, dialects, etc. or a combination of these~\cite{dheram2022toward}.

\subsection{Present work}
Ours is the first work to leverage denoising as a debiasing strategy. In this work, we utilize a previously unexplored combination of signal based and deep learning based denoising techniques to debias ASR systems. Our goal is to reduce performance disparity between male and female speakers without impacting the accuracy of such platforms.

\section{Methodology}
\label{sec:method}

\begin{figure}[!t]
    \centering
    \includegraphics[width=0.45\textwidth,keepaspectratio]{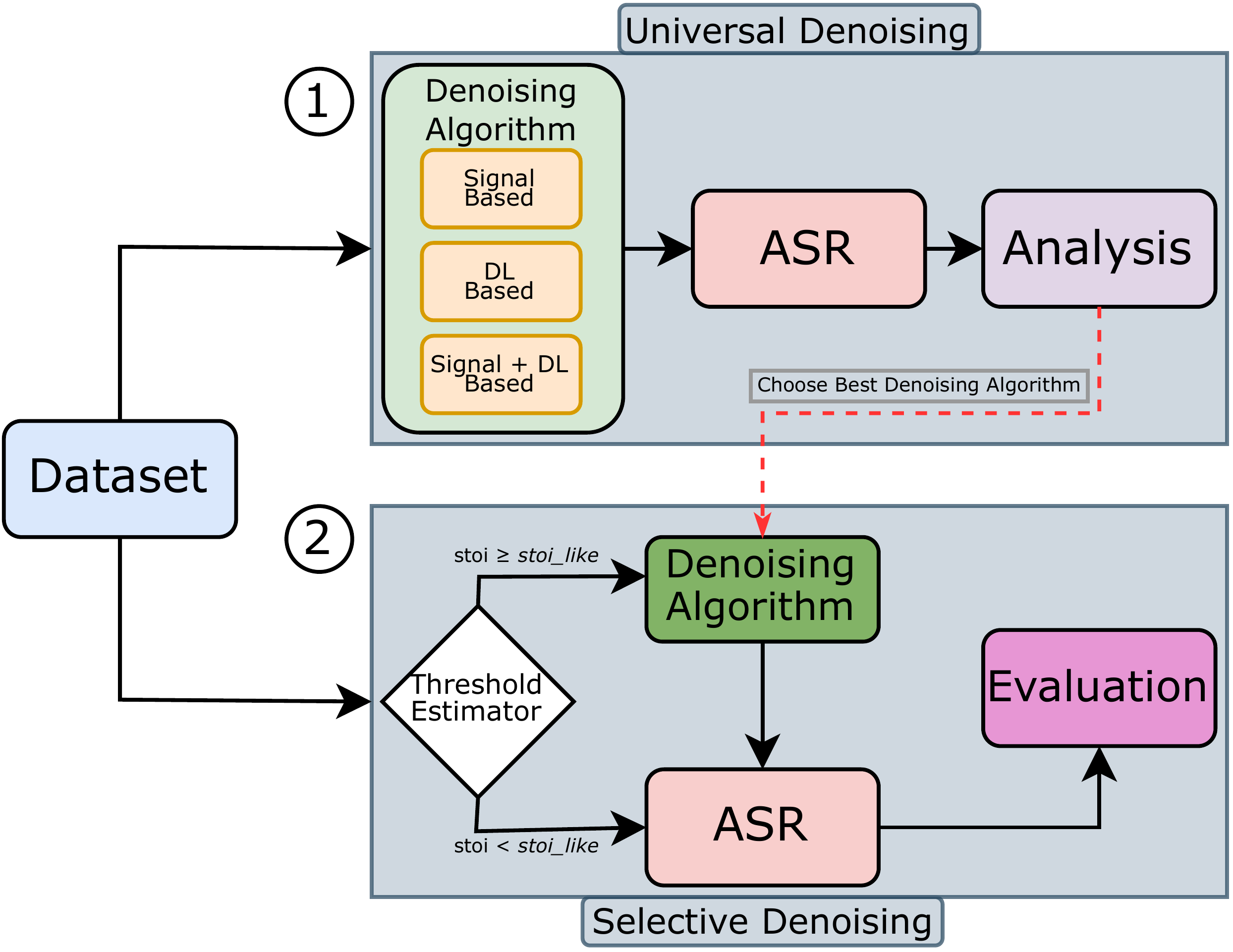}
    \caption{\footnotesize \bf Overview of the \den{} framework for debiasing ASRs.} 
    \label{fig:schematic}
\end{figure}

In this section, we describe the overall \den{} framework (see Figure~\ref{fig:schematic}) for reducing the disparity in the performance of ASRs.

\subsection{Denoising algorithms}
We experiment with both signal based (\sg{} \& \li{}) and deep learning based (\dem{} \& \mg{}) denoising algorithms. A brief description of these algorithms is given below for enhanced readibility.
    \begin{itemize}
        \item  \sg{}: This technique isolates the target speech signal by suppressing background noise based on spectral characteristics. For \sg{}, we use the default parameter settings employed in prior research~\cite{sainburg2020finding}.
        \item  \li{}: This approach aims to enhance the target speech by selectively removing frequencies below a certain threshold. For \li{}, a high-pass filter with a cut-off frequency of 300Hz is used, since most background noise typically falls within this range. Essential speech features, such as harmonics and articulation, generally correspond to frequencies higher than 300Hz~\cite{maccallum2011effects} and are thus effectively retained by the high-pass filter. 
        \item  \dem{}: 
        This is a state-of-the-art neural network architecture specifically designed for speech enhancement tasks, leveraging multichannel convolutional layers to effectively separate speech signals from background noise. In this work we use the open-source version\footnote{https://github.com/facebookresearch/demucs} pre-trained on MUSDB-HQ~\cite{musdb18-hq}.
        \item \mg{}: This approach employs a GAN architecture that mimics human perception of speech quality, learns a noise distribution, and removes it from the signal. In this work we use the Speech Brain Toolkit~\cite{ravanelli2021speechbrain} model\footnote{https://huggingface.co/speechbrain/metricgan-plus-voicebank} pre-trained on Voicebank~\cite{veaux2013voice} recordings sampled at 16kHz (single channel).
    \end{itemize}

    We have also experimented with a combination of the above signal based and deep learning based denoising approaches to evaluate their effectiveness in debiasing ASR performance. 
        
\subsection{Denoising strategy} For denoising the samples using the denoising techniques, following strategies were used:
    \begin{itemize}
        \item \textit{Identification of the best denoising technique}: All speech samples in the dataset are denoised before inference with the ASRs. The technique that performs best in retaining the accuracy while reducing disparity is chosen for selective denoising.
        \item \textit{Selective denoising}: Denoising is selectively applied to samples with STOI (Short-Time Objective Intelligibility) scores below a determined threshold, ensuring that only those with lower perceived speech quality undergo enhancement before ASR inference. The optimal threshold for STOI is obtained using a validation set of 10\% ground-truth transcripts. Grid search on the validation set is used to obtain the threshold value that maximizes disparity reduction while preserving ASR performance. 

    \end{itemize}
  
\subsection{STOI in practice} We have used \textit{stoi\_like} algorithm to estimate STOI. The provided \textit{stoi\_like} algorithm  differs from traditional STOI  calculations in several ways. We summarize the algorithm below for enhanced readability.

In traditional STOI, the intelligibility score is calculated using a clean reference signal $x(t)$ and a degraded test signal $y(t)$. The steps involve aligning the signals, computing their short-time Fourier transforms (STFTs), and then evaluating frame-by-frame correlations. Mathematically, it can be expressed as as follows.

\begin{itemize}
    \item Compute the STFT of the reference signal \( x(t) \) and the degraded signal \( y(t) \):
\[
X(k, n) = \text{STFT}\{x(t)\}
\]
\[
Y(k, n) = \text{STFT}\{y(t)\}
\]
\item Calculate the correlation between the reference and degraded signal frames:
\[
d(x, y) = \sum_{k} \frac{|\langle X(k, n), Y(k, n) \rangle|}{\|X(k, n)\|\|Y(k, n)\|}
\]

\item Map these correlations to predict speech intelligibility.
\end{itemize}

In contrast, the \textit{stoi\_like} algorithm operates on a single degraded audio signal without the need for a reference signal. Instead of using frame-by-frame correlations, this algorithm computes the magnitude of the STFT of the degraded signal and uses mean magnitudes to estimate the energy. It then calculates the noise energy by subtracting the energy of the mean magnitude from the total energy of the signal. The sum of squares of the mean magnitude is computed, and these values are used to derive a metric similar to STOI by taking the logarithm of the ratio of the sum of squares to the noise energy. This approach simplifies the process by removing the necessity of a reference signal and focuses on the signal's internal characteristics, making it potentially more versatile. Mathematically, it can be expressed as as follows.

\begin{itemize}
    \item  Compute the magnitude of the STFT of the degraded signal \( y(t) \):
\[
|Y(k, n)| = \left|\text{STFT}\{y(t)\}\right|
\]

\item Estimate the energy of the signal using mean magnitudes:
\[
E_{\text{signal}} = \sum_{k} \left(\text{mean}(|Y(k, n)|)\right)^2
\]

\item Calculate the noise energy by subtracting the energy of the mean magnitude from the total energy of the signal:
\[
E_{\text{noise}} = \sum_{k} |Y(k, n)|^2 - E_{\text{signal}}
\]

\item Derive a metric similar to STOI by taking the logarithm of the ratio of the sum of squares of the mean magnitude to the noise energy:
\[
\text{stoi\_like} = \log \left(\frac{E_{\text{signal}}}{E_{\text{noise}}}\right)
\]
\end{itemize}

\subsection{Disparity evaluation}
We evaluate the overall performance for an ASR system using the popular \textit{word error rate} (WER) metric\footnote{https://en.wikipedia.org/wiki/Word\_error\_rate}. Before calculating the WER, we standardize the reference and ASR-generated transcripts using procedures such as text normalization, converting numerical expressions to their corresponding textual forms, filler words removal, etc. To evaluate the disparity in performance of the ASR systems for the gender attribute, we utilize the absolute word error gap (AWG) defined as the absolute difference between the median word error rate of the female data points and the male data points. Thus,
\[
\text{AWG} = \left| median(\text{WER}_{\text{f}}) - median(\text{WER}_{\text{m}}) \right| 
\]



\section{Datasets and Platform Evaluated}
We now provide a description of the datasets used and platforms evaluated in this study.
\subsection{Datasets}
\subsubsection{\tie{}} We utilized the recently released TIE dataset~\cite{rai2023deep}, which contains approximately 9.8k audio files sourced from the well-known Indian MOOC platform NPTEL~\cite{krishnan2009nptel}. This dataset spans 8740 hours of technical academic content delivered by 332 instructors, representing diverse demographic segments of the Indian population in terms of age, gender, and geographical location. For the evaluation of our framework, we employed a novel sampling approach with the TIE dataset. Instead of using the entire 50-minute audio files, we processed 3-4 consecutive segments together, resulting in segments of approximately 30 seconds each. This method ensures parity with the default input file size used for speech samples from the other (following) datasets.
\subsubsection{\vox{}} The \vox{} dataset released by Meta, also part of our study, is renowned for its extensive compilation of political speech recordings from the European Parliament. This dataset encompasses thousands of hours of multilingual speech, reflecting a diverse array of languages and formal speech content from parliamentary sessions. In this work, we have used train and test set of speech samples and transcripts in the English language, hosted on HuggingFace\footnote{\url{https://huggingface.co/datasets/facebook/voxpopuli}}.
\subsubsection{\ted{}} The \ted{} dataset offers a comprehensive collection of audio files from TED Talks. This dataset includes a wide range of speakers and topics, capturing various speaking styles and accents from numerous public speaking events. The train and test set of speech data samples and transcripts of this dataset used in our experiments, have been sourced from HuggingFace\footnote{\url{https://huggingface.co/datasets/LIUM/tedlium}}.
\subsubsection{\fle{}} The \fle{} dataset released by Google provides a rich array of speech samples from numerous languages and speakers worldwide. This dataset features extensive coverage of multiple languages and accents, capturing the linguistic diversity of global speech patterns. The English speech train and test set data samples and transcripts of this dataset used for our experiments have been sourced from HuggingFace\footnote{\url{https://huggingface.co/datasets/google/fleurs}}.

The distributional statistics for the aforementioned datasets used in our work are detailed in Table \ref{tab:dataset}.

\begin{table}[!t]
    \scriptsize
    \centering
    \caption{Descriptive Statistics of the datasets used in our study.}
    \begin{tabular}{l r r r r}
    \toprule
    \multicolumn{1}{l}{\textbf{Parameter}} & \textbf{Sampled \tie{}} & \textbf{\vox{}} & \textbf{\ted{}} & \textbf{\fle{}} \\ \midrule
    \#Speakers               &  332        &   774       & 298         & 1510         \\ 
    \#Samples                & 9860        & 9440        & 22370        & 3640        \\ 
     Male (\%)                & 94.2        & 57.0        & 67.8        & 60.3        \\
     Female (\%)               & 5.8        & 43.0        & 32.2        & 39.7        \\
    Avg. duration                 & 24.5 s      & 11.8 s      & 9.4 s      & 19.6 s      \\ 
    Total duration            & 82.4 hrs          & 31.5 hrs       & 58.4 hrs        & 19.2 hrs       \\ 
    \#words & 0.64 M  & 0.23 M  & 0.45 M  & 0.14 M  \\ \bottomrule
    \end{tabular}
    \label{tab:dataset}
\end{table}


\subsection{Open source platforms}
In this work, we study two state-of-the-art open-source ASR platforms -- OpenAI \wh{}~\cite{radford2022robust} and  NVIDIA \nm{}~\cite{kuchaiev2019nemo} for their performance and responsiveness toward our denoising based bias mitigation strategies.

\subsubsection{\wh{}} This platform~\cite{radford2022robust}, released in 2022 by OpenAI, has an end-to-end encoder-decoder transformer and is trained on 680k hours of multilingual and multitask supervised data collected from the web. The model is available in various sizes ranging from tiny (39M parameters) to large (1.5B parameters). In this study, we use the \wh{}-base model\footnote{https://github.com/openai/whisper} having 74M parameters.

\subsubsection{\nm{}} This platform~\cite{kuchaiev2019nemo} provides models for multiple applications -- text processing, speech recognition, and text-to-speech (and vice-versa) processing. 
In this study, we use the \textit{stt\_en\_conformer\_ctc\_large}\footnote{https://huggingface.co/nvidia/stt\_en\_conformer\_ctc\_large} model, which uses the conformer architecture -- a hybrid of CNNs and transformers designed to capture both local and global dependencies in audio data. This specific model is trained on extensive datasets, including thousands of hours of diverse speech data, allowing it to achieve high accuracy in speech-to-text tasks. 

\noindent\textit{System specifications}: We run all our experiments on a GPU server with 16 GB RAM, an NVIDIA T4 GPU with 15 GB memory, and an Intel Xeon processor.




\section{Results}

In this section, we evaluate the \den{} framework. There are two steps in the evaluation. In the first step we identify the best denoising technique in terms of the average WER (AWER) and the average AWG (AAWG). In the second step we use selective denoising \textit{stoi\_like} to further improve the results.

\subsection{Best denoising technique} We wanted to identify the best denoising technique across all the datasets. Hence we mixed the test audio samples from all the datasets to prepare a common set and applied all the different denoising techniques. We then micro-averaged the WER (AWER) and the AWG (AAWG) values across all the datasets. The results are noted in Table~\ref{tab:denoising_algos}. The key observations from this table are listed below.

\begin{itemize}
    \item Denoising strategies generally reduce the disparity in performance over noisy samples, but the extent of improvement varies depending the ASR model and the denoising algorithm used.
    \item Among the two signal processing based denoising techniques used, \li{} provides a notable improvement in AAWG, reducing it from 2.28 to 1.81 for \wh{} and from 2.68 to 2.38 for \nm{}, compared to the noisy baseline across datasets. The overall performance in terms of AWER is also least disturbed for \li{}. 

    \item Of the two deep learning based methods, \dem{} and \mg{}, \dem{} achieves a substantial improvement in AAWG, reducing it from 2.28 to 1.74 for \wh{}. Surprisingly, the \mg{} results in a increase in AAWG for both models compared to the noisy samples. The overall performance in terms of AWER is also least disturbed for \dem{}. 

    \item Since the \li{} and the \dem{} techniques are the best among the signal processing and deep learning approaches respectively, a natural extension to get the benefits of both is to combine them. The combination can be done in two ways -- \dem{} followed by \li{} and vice versa. We find that \dem{} followed by \li{} gives us the best results reducing the AAWG to 1.71 from 2.28 for \wh{} and 2.28 from 2.68 for \nm{}. The overall performance of the two ASR systems in terms of AWER remains largely undisturbed (in fact gets improved for \nm{}). 

\end{itemize}

\begin{table}[!t]
    \begin{center}
    \caption{Performance of denoising algorithms on the mixture of all datasets. \dem{} followed by \li{} is the best denoising technique across the datasets. The best results are highlighted in \colorbox{green!30}{\textbf{boldface}}.}
    \begin{tabular}{lcccc}
    \toprule
    \multicolumn{1}{c}{\multirow{2}{*}{\textbf{\begin{tabular}[c]{@{}c@{}}Denoising\\ strategy\end{tabular}}}} & \multicolumn{2}{c}{\wh{}} & \multicolumn{2}{c}{\nm{}} \\ \cmidrule(lr){2-3} \cmidrule(lr){4-5} 
     & \textbf{AWER} & \textbf{AAWG} & \textbf{AWER} & \textbf{AAWG} \\ \midrule
    Noisy & 9.58           & 2.28          & 12.22          & 2.68         \\ \hline
    \sg{}    & 14.20           & 2.10          & 12.86          & 2.84         \\ 
    \li{}    & 10.76           & 1.81          & 12.14          & 2.38         \\ \hline
    \mg{}    & 13.26           & 2.30          & 14.38           & 3.88         \\ 
    \dem{}    & 10.21           & 1.74          & 13.54           & 2.69          \\ \hline
    \li{} + \dem{} & 11.30          & 1.90          & 12.47 & 3.05 \\ 
    \dem{} + \li{} & \cellcolor{green!30}\textbf{10.32} & \cellcolor{green!30}\textbf{1.71} & \cellcolor{green!30}\textbf{11.98}          & \cellcolor{green!30}\textbf{2.28}          \\ \bottomrule
    \end{tabular}
    \label{tab:denoising_algos}
    \end{center}
\end{table}



\subsection{Selective denoising}

In the previous section we identified that \dem{} followed by \li{} turned out to be the best denoising technique across the datasets. However, when we observed the results manually we found that for a group of data points where the speech is unintelligible denoising is effective. On the other hand if the speech is intelligible then denoising has an opposite impact, i.e., it deteriorates the signal quality. Accordingly, we resorted to selective denoising based on an STOI threshold obtained for each dataset using the method described in section~\ref{sec:method}. The results from this approach are noted in Table \ref{tab:selective_denoising}. The key observations from these results are as follows.
\begin{itemize}
    \item Selective denoising dramatically reduces the AWG for all the datasets and both the ASR models.
    \item Among the datasets selective denoising is most effective in case of \vox{} with AWG dropping to 0 for both the ASR models. Other datasets also show steady gains.
    \item Selective denoising is more effective on \nm{} among the two ASR models. 
\end{itemize}

\begin{table}[!t]
    \footnotesize
    \begin{center}
    \caption{The results of the selective denoising approach using \dem{} followed by \li{} for all the individul datasets. The best values are highlighted in \colorbox{green!30}{\textbf{boldface}}}.
    \begin{tabular}{c c cc cc}
    \toprule
    \multirow{2}{*}{\textbf{Model}} & \multirow{2}{*}{\textbf{Dataset}} & \multicolumn{2}{c}{\textbf{No denoising}}        & \multicolumn{2}{c}{\textbf{Selective denoising}} \\ \cmidrule(lr){3-4} \cmidrule(lr){5-6} 
    & & \textbf{WER} & \textbf{AWG} & \textbf{WER} & \textbf{AWG} \\ \midrule
    \multirow{4}{*}{\wh{}}        & Sampled \tie{}                               & 11.63        & 1.59          & 12.06        & \cellcolor{green!30}\textbf{1.41}          \\  
                                    & \vox{}                      & 6.52        & 0.22          & 7.14        & \cellcolor{green!30}\textbf{0.00}          \\  
                                    & \ted{}                       & 2.86        & 0.31          & 3.12         & \cellcolor{green!30}\textbf{0.09}             \\  
                                    & \fle{}                          & 10.52         & 0.09          & 10.71         & \cellcolor{green!30}\textbf{0.00}          \\ \hline
    \multirow{4}{*}{\nm{}}           & Sampled \tie{}                               & 18.60        & 2.68          & 18.64        & \cellcolor{green!30}\textbf{2.08}          \\  
                                    & \vox{}                      & 2.71        & 0.08          & 3.33        & \cellcolor{green!30}\textbf{0.00}         \\  
                                    & \ted{}                        & 5.26         & 1.13          & 5.26         & \cellcolor{green!30}\textbf{0.26}             \\  
                                    & \fle{}                          & 5.48         & 1.12          & 5.55         & \cellcolor{green!30}\textbf{0.88}          \\ \bottomrule
    \end{tabular}
    \label{tab:selective_denoising}
    \end{center}
\end{table}

\if{0}
\subsection{Noisy Samples}
The key takeaways of the ASR inference done using noisy samples from the datasets are as follows:
\begin{itemize}
    \item As per Table \ref{tab:selective_denoising}, for both models, the Sampled TIE dataset has the highest WER and AWG, indicating a significant challenge in addressing gender disparity in ASR performance with this dataset in noisy conditions. 
    \item Also Table \ref{tab:selective_denoising} illustrates that for the other datasets, the WER is comparatively lower than that of the sampled TIE dataset, but gender disparity in terms of AWG persists in the performance of both ASR models.
    \item Table \ref{tab:denoising_algos} shows that for the combined speech samples from the test sets of other datasets and the entire sampled TIE, both AAWER and AAWG are higher for the NVIDIA Nemo model compared to the Whisper model.
\end{itemize}

\subsection{Universally Denoised Samples}
As evident in Table \ref{tab:denoising_algos}, denoising strategies generally reduces the disparity in performance over noisy samples, but the extent of improvement varies depending the ASR model and the denoising algorithm used. Some key takeaways of universal denoising results from Table \ref{tab:denoising_algos} are following:
\begin{itemize}
    \item Among the two signal-based denoising techniques used, LE provides a notable improvement in AAWG, reducing it from 2.28 to 1.81 for Whisper and from 2.68 to 2.38 for NVIDIA Nemo, compared to the noisy baseline across datasets.

    \item Of the two deep learning-based methods, DM and LE, the DM strategy achieves a substantial improvement in AAWG, reducing it from 2.28 to 1.74 for Whisper. In contrast, the MG strategy increases AAWG for both models compared to the noisy samples.

    \item Combining LE and DM due to their superior performance in reducing gender-based disparity compared to their counterparts SG and MG yielded favorable results. Applying DM followed by LE to noisy samples before inference with ASRs resulted in further improvement in AAWG, reducing it from 2.28 to 1.71 for Whisper and from 2.68 to 2.28 for NVIDIA Nemo. However, applying LE followed by DM did not produce similar improvements.

\end{itemize}
\fi
In conclusion, our experimental results based on the \den{} framework highlight the effectiveness of various denoising strategies in reducing gender-based disparity in ASRs across different datasets without hurting the overall performance much. 

\section{Discussion}

In this section, we deep dive into our findings and explore the broader impact of the denoising strategies on ASR performance and gender disparity. By analyzing the results from various datasets and denoising techniques, we aim to provide a comprehensive understanding of how these strategies influence overall accuracy and fairness in ASR systems.

\subsection{Error analysis}



Some of the key observations from the transcript examples generated through baseline models and the models after applying \dem{} + \li{} and selective denoising are enumerated below.

\begin{itemize}
   \item For the base sample of the male voice in the \tie{} dataset, the \wh{} model produces a noisy transcript that seems to hallucinate, generating nonsensical phrases such as `n n n n n n n'. This issue gets resolved when the same base sample is first denoised using \dem{} + \li{} and then transcribed using \wh{}. Selective denoising continues to maintain this result.
    \item One of the hardest cases is the transcription, \textit{ottawa is canada's charming bilingual capital and features an array of art galleries and museums that showcase canada's past}, of the female voice in the \fle{} dataset using the \nm{} model. Neither \dem{} + \li{} denoising nor selective denoising is effective in resolving the transcription error. A possible reason could be the alliteration in the speech segment `canada's charming \dots~capital \dots' that poses a more complex challenge for the model. 
    
\end{itemize}

Overall, the selective denoising method consistently provides transcripts that are more coherent and contextually accurate irrespective of gender compared to the blindly applying \dem{} + \li{} based denoising over the whole dataset. However, none of these methods completely eliminate all errors, particularly in complex speech situations. The ground-truth data highlights the gap that still exists between automated denoising methods and human-level transcription accuracy.

\subsection{Generalizing to other attributes}

The effectiveness of the selective denoising algorithm extends beyond addressing gender disparities in ASR performance to other demographic attributes as well. Table \ref{tab:general_tab} demonstrates the impact of this algorithm for various other demographic attributes present in the TIE dataset, including native region, experience group, and discipline group. The table compares the AWG for both \wh{} and \nm{} models before and after applying selective denoising on speech samples from sampled \tie{} dataset. For all these groups we observe that the disparity is reduced via selective denoising for the \wh{} model. For the \nm{} model the disparity is reduced in two out of three demographic groups through selective denoising. This shows that the \den{} framework is significantly powerful and can be seamlessly used for disparity reduction for any other demographic attributes.


\begin{table}[!t]
    \begin{center}
    \caption{Debiasing impact of selective denosing algorithms on other demographic categories of Sampled \tie{} dataset.  The best results are highlighted in \colorbox{green!30}{\textbf{boldface}}. AWG (B): AWG for baseline, AWG (D): AWG after selective denoising.}
    \begin{tabular}{lcccc}
    \toprule
    \multicolumn{1}{c}{\multirow{2}{*}{\textbf{\begin{tabular}[c]{@{}c@{}} Annotated\\ category\end{tabular}}}} & \multicolumn{2}{c}{\wh{}} & \multicolumn{2}{c}{\nm{}} \\ \cmidrule(lr){2-3} \cmidrule(lr){4-5} 
     & \textbf{AWG (B)} & \textbf{AWG (D)} & \textbf{AWG (B)} & \textbf{AWG (D)} \\ \midrule
    Native region &  2.14          & \cellcolor{green!30}\textbf{2.09}          & 1.17          & 1.20         \\ 
    Experience group & 3.75         & \cellcolor{green!30}\textbf{3.55}        & 2.20          & \cellcolor{green!30}\textbf{2.14}        \\ 
    Discipline group & 2.59           & \cellcolor{green!30}\textbf{2.08}          & 0.32          & \cellcolor{green!30}\textbf{0.23}         
             \\ \bottomrule
    \end{tabular}
    \label{tab:general_tab}
    \end{center}
\end{table}


\if{0}\subsection{Effectiveness of Denoising Techniques}
Although signal-based and DL-based denoising methods share the common objective of identifying and segregating noise components from speech signals to enhance overall intelligibility and Signal to Noise Ratio (SNR), their effectiveness can vary depending on the dataset and ASR model. However, the impact becomes more broadly applicable, particularly in reducing gender disparity, when both signal and DL-based methods (DM + LE) are simultaneously employed for universal denoising of datasets before ASR inference, as opposed to individually applying the denoising algorithms.

Using both signal-based and DL-based denoising methods simultaneously (DM + LE) for either universal denoising or selectively denoising speech samples before ASR inference enhances denoising effectiveness by leveraging the strengths of each approach, resulting in more consistent and impactful reduction of gender disparity in ASR performance.

\subsection{Debiasing Impact of Selective Denoising}\fi

\subsection{Intelligibility threshold analysis}

\begin{figure}[!t]
   
    \begin{subfigure}[b]{\textwidth}
        \includegraphics[width=0.45\textwidth,keepaspectratio]{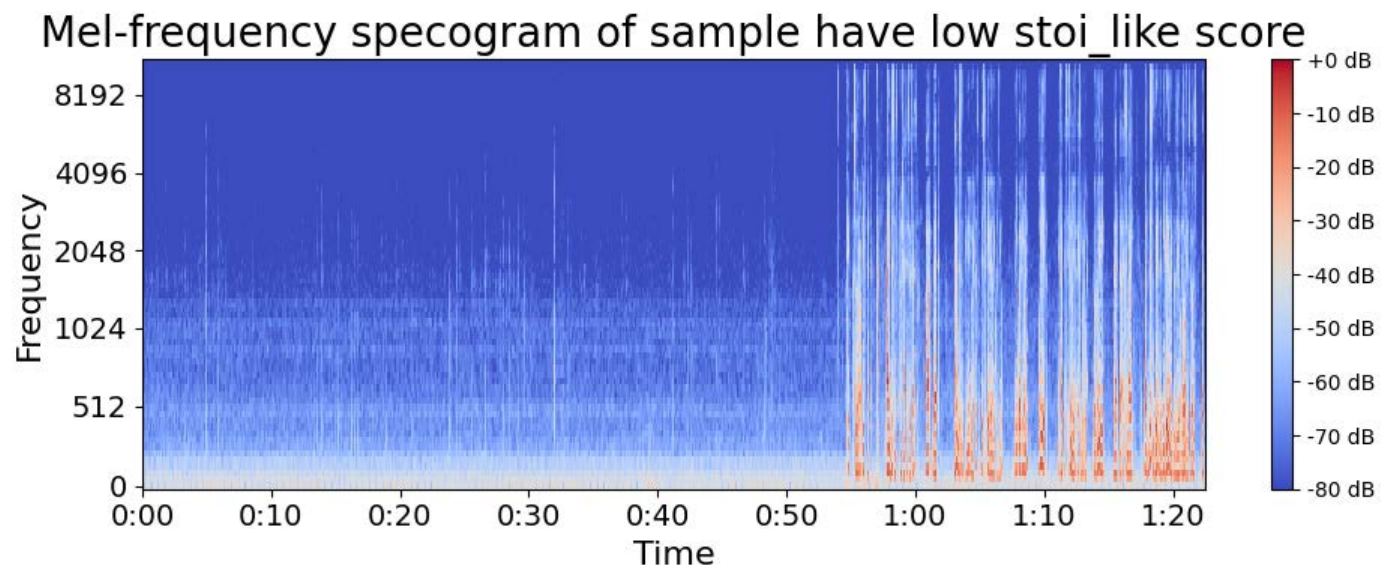}
        \label{fig:image1}
    \end{subfigure}
    \vfill
    \begin{subfigure}[b]{\textwidth}
       
        \includegraphics[width=0.45\textwidth,keepaspectratio]{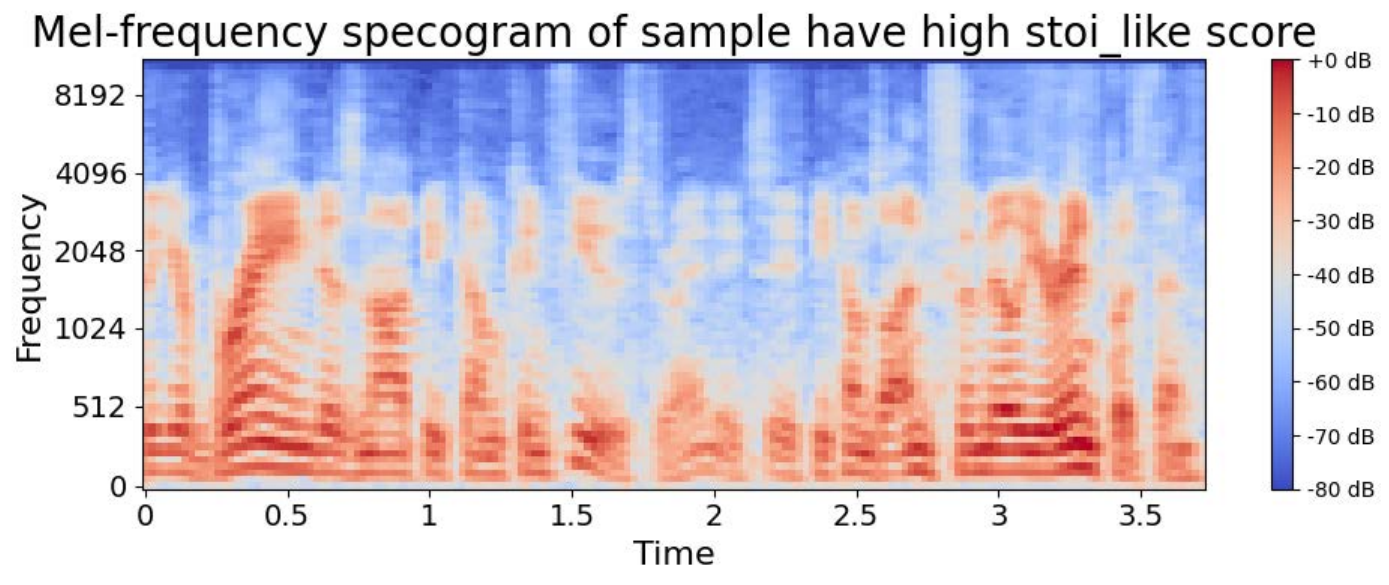}
        \label{fig:image2}
    \end{subfigure}
    \caption{(\textbf{Top}) Spectrogram of speech sample with low \textit{stoi\_like} score having significant noise in lower frequencies. (\textbf{Bottom}) Spectrogram of speech sample with high \textit{stoi\_like} score having more noise in higher frequencies.}
    \label{fig:two_images}
\end{figure}

The spectrogram of the speech sample in Figure \ref{fig:two_images} (\textbf{Top}) with a low \textit{stoi\_like} score of -1.51 reveals significant noise across multiple time segments, severely impacting the intelligibility of the audio. This degradation is evident in the performance of the two ASR models, \wh{} and \nm{}. While the ground-truth transcript is \textit{``so the properties that the hamming distance function satisfies or that the hamming distance between any two vectors is strictly greater than or equal to zero with equality holding only if in fact if and only if x equal to y because the only way that x and y the distance greater than being greater than or equal to zero it is obvious because we are''}, \wh{}, outputs only a single word \textit{``so,''} indicating a failure to process the noisy input effectively. In contrast, \nm{} performs markedly better, producing a more coherent transcription, \textit{``so the properties that the humming distance function satisfies are that the humming distance between any two vectors is strictly greater than equal to zero with equality holding only if in fact it is if and only if x equal to y because the only way that x and y the distance greater than being greater than equal to zero is obvious because''} that closely follows the structure and content of the ground truth, with slight inaccuracies. This comparison underscores the challenges faced by ASR systems in handling noisy inputs and hence, as stated earlier, there is a difference in threshold values between the ASR models for the same dataset.

On the other hand, the mel-frequency spectrogram in Figure \ref{fig:two_images} (\textbf{Bottom}) has a high \textit{stoi\_like} score of 12.35 due to the distinct and consistent formant structures, clear temporal patterns, and significant amplitude variations. The transcription generated by both ASR models \wh{} and \nm{} corresponds exactly to the ground-truth transcript, \textit{``so your brain does not become confused it becomes unfamiliar input then''}. This indicates the presence of noise in speech samples does not warrant need of denoising for each sample and hence our strategy of selective denoising based on intelligibility threshold works better in the task of disparity mitigation.

\section{Conclusion}

In this study, we have investigated the effectiveness of the \den{} framework in mitigating the transcription errors in ASR systems, with a focus on addressing gender disparity. We applied both signal based and deep learning based denoising techniques on multiple public datasets viz. \tie{}, \vox{}, \ted{} and \fle{} and used two very popular open-source ASRs -- \wh{} and \nm{} for transcript generation. Using an intelligibility based selective denoising we obtain substantial improvements in transcription accuracy for both male and female speakers across all the datasets. Our analysis revealed that the \den{} framework not only enhances ASR performance for gender-specific attributes but also extends its benefits to other demographic categories, such as native region, experience group, and discipline group, as evidenced by our results with the \tie{} dataset.

Overall, the \den{} framework shows significant promise in advancing ASR systems by enhancing their robustness and fairness across diverse user demographics. This method can be seamlessly integrated into real-time ASRs at the pre-processing stage, operating as a plug-and-play solution without requiring fine-tuning of the ASR itself. Future work should continue to refine these algorithms and explore complementary methods to address the challenges of achieving truly inclusive and accurate ASR systems for all users.

\bibliographystyle{splncs04}
\bibliography{reference}

\vspace{12pt}

\end{document}